\documentclass[12pt]{iopart}

\begin{document}

%old title
%\title[Influence of compressibility on advection of  a passive scalar]
%{Influence of compressibility on the scaling regimes of a passive
%scalar advected by turbulent velocity field with finite correlation
%time}

% new title
\title[Compressible advection of a passive scalar]
{Compressible advection of a passive scalar: Two-loop scaling
regimes}

\author{M Hnati\v{c}$^{1,2}$,
E Jur\v{c}i\v{s}inov\'a$^{1}$\footnote{Present address: Laboratory
of Information Technologies, JINR, 141 980 Dubna, Russia}, M
Jur\v{c}i\v{s}in$^{1}$\footnote{Present address: Laboratory of
Theoretical Physics, JINR, 141 980 Dubna, Russia} and M
Repa\v{s}an$^{1}$}

\address{$^{1}$ Institute of Experimental Physics, Slovak Academy of Sciences,
                Watsonova 47, 040 01, Ko\v{s}ice, Slovakia}
\address{$^{2}$  Department of Mathematics Faculty of Civil Engineering, Technical University
Vysoko\v{s}kolsk\'a 4, 040 01 Ko\v{s}ice, Slovakia}
\eads{\mailto{hnatic@saske.sk}, \mailto{eva.jurcisinova@post.sk},
\mailto{jurcisin@thsun1.jinr.ru}, \mailto{repasan@saske.sk}}
\begin{abstract}
The influence of compressibility on the stability of the scaling
regimes of the passive scalar advected by a Gaussian
% solenoidal
velocity field with finite correlation time is investigated by the
field theoretic renormalization group within two-loop approximation.
The influence of compressibility on the scaling regimes is discussed
as a function of the exponents $\varepsilon$ and  $\eta$, where
$\varepsilon$ characterizes the energy spectrum of the velocity
field in the inertial range
% $E\propto k^{1-\varepsilon}$
$E\propto k^{1-2\varepsilon}$, and $\eta$ is related to the
correlation time at the wave number $k$ which is scaled as
$k^{-2+\eta}$. The restrictions given by nonzero compressibility  on
the regions with stable infrared fixed points which correspond to
the stable infrared scaling regimes are discussed in detail.
% added text follows
A special attention is paid to the case of so-called frozen velocity
field, when the velocity correlator is time independent. In this
case, explicit inequalities which must be fulfilled in the plane
$\varepsilon-\eta$ are determined within two-loop approximation. The
existence of a "critical" value $\alpha_c$ of the parameter of
compressibility $\alpha$ at which one of the two-loop conditions is
canceled as a result of the competition between compressible and
incompressible terms is discussed. Brief general analysis of the
stability of the scaling regime of the model with finite
correlations in time of the velocity field within two-loop
approximation is also given.
\end{abstract}
\submitto{\JPA}


%\maketitle

\section{Introduction}\label{sec1}

One of the main problems in the modern theory of fully developed
turbulence is to verify the validity of the basic principles of
Kolmogorov-Obukhov (KO) phenomenological theory and their
consequences within the framework of a microscopic model
\cite{MonYag75,Frisch95}. On the other hand, recent experimental,
numerical and theoretical studies signify the existence of
deviations from the well-known Kolmogorov scaling behavior. The
scaling behavior  of the velocity fluctuations with exponents, which
values are different from Kolmogorov ones, is known as anomalous and
is associated with intermittency phenomenon \cite{Frisch95}. Even
thought the understanding of the intermittency and anomalous scaling
within the theoretical description of the fluid turbulence on basis
of the "first principles", i.e., on the stochastic Navier-Stokes
equation, still remains an open problem, considerable progress has
been achieved in the studies of the simplified model systems which
share some important properties of the real turbulence.

The crucial role in these studies is played by models of advected
passive scalar field \cite{Obu49}. Maybe the most known model of
this type is a simple model of a passive scalar quantity advected by
a random Gaussian velocity field, white in time and self-similar in
space, the so-called Kraichnan's rapid-change model \cite{Kra68}. It
was shown by both natural and numerical experimental investigations
that the deviations from the predictions of the classical KO
phenomenological theory is even more strongly displayed for a
passively advected scalar field than for the velocity field itself
(see, e.g., \cite{FaGaVe01} and references cited therein). At the
same time, the problem of passive advection is much more easier to
be consider from theoretical point of view. There, for the first
time, the anomalous scaling was established on the basis of a
microscopic model \cite{Kraichnan94}, and corresponding anomalous
exponents was calculated within controlled approximations (see
review \cite{FaGaVe01} and references therein).

In paper \cite{AdAnVa98+} the field theoretic renormalization group
(RG) and operator-product expansion (OPE) were used in the
systematic investigation of the rapid-change model. It was shown
that within the field theoretic approach the anomalous scaling is
related to the very existence of so-called "dangerous" composite
operators with negative critical dimensions in OPE (see, e.g.,
\cite{Vasiliev,AdAnVa99} for details). Important advantages of the
RG approach are its universality and calculational efficiency: a
regular systematic perturbation expansion for the anomalous
exponents was constructed, similar to the well-known
$\epsilon$-expansion in the theory of phase transitions.

Afterwards, various generalized descendants of the Kraichnan model,
namely, models with inclusion of large and small scale anisotropy
\cite{AdAnHnNo00}, compressibility \cite{AdAn98} and finite
correlation time of the velocity field \cite{Antonov99,Antonov00}
were studied by the field theoretic approach. General conclusion is:
the anomalous scaling, which is the most important feature of the
Kraichnan rapid change model, remains valid for all generalized
models.

In paper \cite{Antonov99} the problem of a passive scalar advected
by the Gaussian self-similar velocity field with finite correlation
time \cite{all2} was studied by the field theoretic RG method.
There, the systematic study of the possible scaling regimes and
anomalous behavior was present at one-loop level. The two-loop
corrections to the anomalous exponents were obtained in
\cite{AdAnHo02}. In paper \cite{Antonov00} the influence of
compressibility on the problem studied in \cite{Antonov99} was
analyzed. In what follows, we shall continue with the investigation
of this model from the point of view of the influence of
compressibility on the stability of the scaling regimes within
two-loop approximation.
% added text follows
It can lead to sufficient restrictions of the parameter space where
the stable fixed points can exist. This, as we shall see rather
complicated task, is the first nontrivial step on the way to
understand the influence of the compressibility of the system on the
two-loop corrections to anomalous dimensions of the measurable
quantities \cite{AdAnHo02}.

\section{Description of the model}\label{sec2}

We consider the advection of a passive scalar field $\theta \equiv
\theta(x)\equiv \theta(t, {\bf x})$ which is described by the
stochastic equation
\begin{equation}
\partial_t \theta + v_i \partial_i \theta=\nu_0 \Delta
\theta + f^{\theta},\label{scalar1}
\end{equation}
where $\partial_t \equiv \partial/\partial t$, $\partial_i \equiv
\partial/\partial x_i$, $\nu_0$ is the coefficient of molecular
diffusivity (hereafter all parameters with a subscript $0$ denote
bare parameters of unrenormalized theory; see below), $\Delta \equiv
\partial^2$ is the Laplace operator, and $f^{\theta}
\equiv f^{\theta}(x)$ is a Gaussian random noise with zero mean and
correlation function
\begin{equation}
\langle f^{\theta}(x) f^{\theta}(x^{\prime})\rangle =
\delta(t-t^{\prime})C({\bf r}/L), \,\,\, {\bf r}={\bf x}-{\bf
x^{\prime}},\label{correlator}
\end{equation}
where parentheses $\langle...\rangle$ hereafter denote average over
corresponding statistical ensemble. The noise maintains the
steady-state of the system but the concrete form of the correlator
is not essential. The only condition which must be fulfilled by the
function $C({\bf r}/L)$ is that it must decrease rapidly for
$r\equiv |{\bf r}| \gg L$, where $L$ denotes an integral scale
related to the stirring. The velocity field ${\bf v(x)}$ obeys a
Gaussian distribution with zero mean and correlator
% the reference is added
\cite{Antonov00}
\begin{eqnarray}
%\hspace{-2.1cm}
\fl \langle v_i(x) v_j(x^{\prime}) \rangle &=&
D^v_{ij}(x,x^{\prime})\label{corv} \\ &=&\int \frac{d\omega d^d
k}{(2\pi)^{d+1}} \left(P_{ij}({\bf k}) + \alpha Q_{ij}({\bf k})
\right) \tilde{D}^v(\omega,k) \exp[-i\omega(t-t^{\prime})+i{\bf
k}({\bf x}-{\bf x^{\prime}})],\nonumber
\end{eqnarray}
where $k=|{\bf k}|$ is the wave number, $\omega$ is frequency, $d$
is the dimensionality of the ${\bf x}$ space. In what follows, we
shall work with compressible velocity field which is demonstrated by
the form of the tensor structure of the correlator (\ref{corv}),
namely, it consists of two parts: the standard transverse projector
$P_{ij}({\bf k})=\delta_{ij}-k_ik_j/k^2$, and the longitudinal
projector $Q_{ij}({\bf k})=k_i k_j/k^2$ which is related to
compressibility. The parameter $\alpha \geq 0$ is a free parameter.
The value $\alpha=0$ corresponds to the divergence-free
(incompressible) advecting velocity field. The function
$\tilde{D}^v$ is chosen as follows \cite{Antonov99,Antonov00}
\begin{equation}
\tilde{D}^v(\omega, k) = \frac{g_0 \nu_0^3
k^{4-d-2\varepsilon-\eta}}{(i\omega+u_0 \nu_0
k^{2-\eta})(-i\omega+u_0 \nu_0 k^{2-\eta})}.\label{corrvelo}
\end{equation}
The correlator (\ref{corrvelo}) is related to the energy spectrum
via the frequency integral
\begin{equation}
E(k)\simeq k^{d-1} \int d\omega \tilde{D}^v(\omega, k) \simeq
\frac{g_0 \nu_0^2}{u_0} k^{1-2\varepsilon}.
\end{equation}
It means that the coupling constant $g_0$
% beginning of added text
(more precisely $g_0/u_0$ \cite{Antonov00})
% end of the added text
and the exponent $\varepsilon$ describe the equal-time velocity
correlator or, equi\-va\-lent\-ly, energy spectrum. On the other
hand, the constant $u_0$ and the second exponent $\eta$ are related
to the frequency $\omega \simeq u_0 \nu_0 k^{2-\eta}$ which
characterizes the mode $k$. Thus, in our notation, the value
$\varepsilon=4/3$ corresponds to the well-known Kolmogorov
"five-thirds law" for the spatial statistics of velocity field, and
$\eta=4/3$ corresponds to the Kolmogorov frequency. Simple
dimensional ana\-ly\-sis shows that the charges $g_0$ and $u_0$ are
related to the characteristic ultraviolet (UV) momentum scale
$\Lambda$ (of the order of inverse Kolmogorov length) by
\begin{equation}
g_0\simeq \Lambda^{2\varepsilon + \eta},\,\,\, u_0\simeq
\Lambda^{\eta}.
\end{equation}

In the end of this section, let us briefly discuss two important
limits of the considered model (\ref{corv}), (\ref{corrvelo}) (see
also \cite{Antonov99,Antonov00}). First of them is so-called
rapid-change model limit when $u_0\rightarrow \infty$ and
$g_0^{\prime}\equiv g_0/u_0^2=$ const,
\begin{equation}
\tilde{D}^v(\omega, k)\rightarrow g_0^{\prime} \nu_0
k^{-d-2\varepsilon + \eta},
\end{equation}
and the second one is so-called quenched (time-independent or
frozen) velocity field limit which is defined by $u_0\rightarrow 0$
and $g_0^{\prime\prime}\equiv g_0/u_0=$ const,
\begin{equation}
\tilde{D}^v(\omega, k)\rightarrow g_0^{\prime\prime} \nu_0^2
k^{-d+2-2\varepsilon} \pi \delta(\omega).
\end{equation}
Here the velocity correlator is independent of time in the $t$
representation.

\section{Field Theoretic Formulation of the Model}\label{sec3}

The stochastic problem (\ref{scalar1})-(\ref{corv}) is equivalent to
the field theoretic model of the set of fields $\Phi \equiv
\{\theta, \theta^{\prime}, {\bf v}\}$ (see, e.g.,
\cite{Vasiliev,ZinnJustin}) with action functional
\begin{eqnarray}
\hspace{-1cm} S(\Phi)=&-&\frac{1}{2} \int dt_1\,d^d{\bf
x_1}\,dt_2\,d^d{\bf x_2} \,\,v_i(t_1,{\bf x_1}) [D^{v}_{ij}(t_1,{\bf
x_1};t_2,{\bf x_2})]^{-1} v_j(t_2,{\bf x_2})  \nonumber \\
&+& \int dt\,d^d{\bf x}\,\, \theta^{\prime}\left[-\partial_t \theta
- v_i\partial_i\theta+\nu_0\triangle\theta \right], \label{action1}
\end{eqnarray}
where, in what follows, unimportant term related to the noise
(\ref{correlator}) is omitted, $\theta^{\prime}$  is an auxiliary
scalar field, and summations are implied over the vector indices.
The second line in (\ref{action1}) represent the Martin-Siggia-Rose
action for the stochastic problem (\ref{scalar1}) at fixed velocity
field ${\bf v}$, and the first line describes the Gaussian averaging
over ${\bf v}$ defined by the correlator $D^v$ in (\ref{corv}) and
(\ref{corrvelo}).

Standardly, the formulation through the action functional
(\ref{action1}) replaces the statistical averages of random
quantities in the stochastic problem (\ref{scalar1})-(\ref{corv})
with equi\-va\-lent functional averages with weight $\exp S(\Phi)$.
Ge\-ne\-ra\-ting functionals of total Green functions G(A) and
connected Green functions W(A) are then defined by the functional
integral
\begin{equation}
G(A)=e^{W(A)}=\int {\cal D}\Phi \,\, e^{S(\Phi) +
A\Phi},\label{green}
\end{equation}
where $A(x)=\{A^{\theta},A^{\theta^{\prime}},{\bf A^{v}}\}$
represents a set of arbitrary sources for the set of fields $\Phi$,
${\cal D}\Phi \equiv {\cal D}\theta{\cal D}\theta^{\prime}{\cal
D}{\bf v}$ denotes the measure of functional integration, and linear
form $A\Phi$ is defined as
\begin{equation}
A\Phi= \int d\,x
[A^{\theta}(x)\theta(x)+A^{\theta^{\prime}}(x)\theta^{\prime}(x) +
A_i^{v}(x) v_i(x)].\label{form}
\end{equation}

\input epsf
   \begin{figure}[t]
     \vspace{1cm}
       \begin{center}
       \leavevmode
       \epsfxsize=5cm
       \epsffile{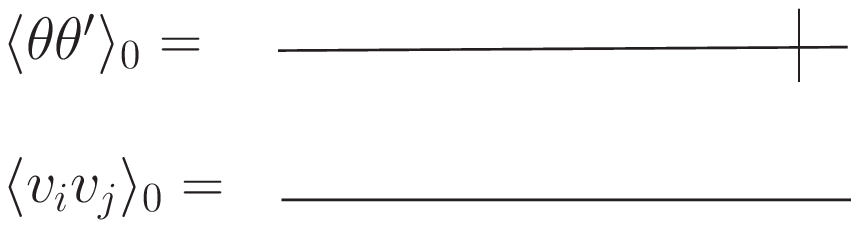}
   \end{center}
\vspace{0cm} \caption{The graphical representation of the
propagators of the model.\label{propagators}}
\end{figure}

\input epsf
   \begin{figure}[t]
     \vspace{0cm}
       \begin{center}
       \leavevmode
       \epsfxsize=5cm
       \epsffile{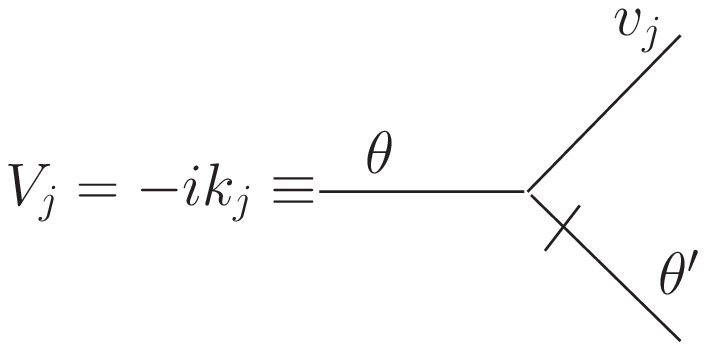}
   \end{center}
\vspace{0cm} \caption{The interaction vertex of the model
\\  (wave-number-frequency representation). \label{vertex}}
\end{figure}

Action (\ref{action1}) is given in a form convenient for a
rea\-li\-za\-tion of the field theoretic perturbation analysis with
the standard Feynman diagrammatic technique. The matrix of bare
propagators is obtained from the quadratic part of the action. The
wave-number-frequency representation of, in what follows, important
propagators are: a) the bare propagator $\langle\theta
\theta^{\prime}\rangle_0$ defined as
\begin{equation}
\langle\theta \theta^{\prime}\rangle_0=\langle\theta^{\prime}
\theta\rangle^*_0=\frac{1}{-i\omega+\nu_0 k^2},
\end{equation}
and b) the bare propagator for the velocity field $\langle v
v\rangle_0$ given directly by (\ref{corrvelo}), namely
\begin{equation}
\langle v_i v_j\rangle_0 = \left(P_{ij}({\bf k}) + \alpha
Q_{ij}({\bf k}) \right) D^v(\omega, k).
\end{equation}
Their graphical representation is present in figure
\ref{propagators}.

The triple (interaction) vertex $-\theta^{\prime} v_j\partial_j
\theta = \theta^{\prime} v_j V_j \theta $ is present in figure
\ref{vertex}, where momentum ${\bf k}$ is flowing into the vertex
via the
% auxiliary is change to
scalar
field $\theta$.

\section{Renormalization and RG analysis}\label{sec4}

The model under consideration is logarithmic at $\varepsilon=\eta=0$
(the coupling constants $g_0$, and $u_0$ are dimensionless),
therefore the UV divergences in the correlation functions have the
form of the poles in $\varepsilon, \eta$, and their linear
combinations.

% the old text:
%The quantity which plays a central role in the renormalization of
%the model, namely, the role of the formal index of the UV
%divergence, is the total canonical dimension of an arbitrary
%one-irreducible correlation (Green) function $\Gamma=\langle \Phi
%\cdots \Phi \rangle_{1-ir}$. It is given as follows
%\begin{equation}
%d_{\Gamma}=d^k_{\Gamma}+2 d^{\omega}_{\Gamma}=d+2-N_{\Phi} d_{\Phi},
%\end{equation}
%where $N_{\Phi}=\{N_{\theta},N_{\theta^{\prime}},N_{{\bf v}}\}$ are
%the numbers of corresponding fields entering into the function
%$\Gamma$, and summation over all types of fields is implied. It is
%well-known that superficial UV divergences, whose removal requires
%counterterms, can be presented only in those Green functions
%$\Gamma$ for which the total canonical index $d_{\Gamma}$ is
%non-negative integer.

% the beginning of the new text
The crucial role in the renormalization of the model is played by
the total canonical dimension of an arbitrary one-particle
irreducible correlation (Green) function $\Gamma=\langle \Phi \cdots
\Phi \rangle_{1-ir}$. It plays the role of the formal index of the
UV divergence and it is given as follows \cite{Vasiliev,AdAnVa99}
\begin{equation}
d_{\Gamma}=d^k_{\Gamma}+2 d^{\omega}_{\Gamma}=d+2-N_{\Phi} d_{\Phi},
\end{equation}
where $N_{\Phi}=\{N_{\theta},N_{\theta^{\prime}},N_{{\bf v}}\}$ are
the numbers of corresponding fields entering into the function
$\Gamma$,
% added text follows
$d^k_{\Gamma}$ and $d^{\omega}_{\Gamma}$ are the canonical momentum
dimension and the canonical frequency dimension of the function
$\Gamma$, respectively,
% end of the added text
and summation over all types of fields is implied. In what follows,
we shall use the definitions of the canonical dimensions of the
fields $\Phi$ as they are given in \cite{Antonov99,Antonov00}. It is
well-known that superficial UV divergences, whose removal requires
counterterms, can be presented only in those Green functions
$\Gamma$ for which the total canonical index $d_{\Gamma}$ is
non-negative integer.
% the end of the new text

From the dimensional analysis of the model (see, e.g.,
\cite{Antonov99,Antonov00}), we conclude that for any $d$,
superficial UV divergences can exist only in the 1-irreducible
functions $\langle \theta^{\prime} \theta \rangle_{1-ir}$ and
$\langle \theta^{\prime} \theta  {\bf v}\rangle_{1-ir}$. To remove
them one needs to include into the action functional the counterterm
of the form $\theta^{\prime} \triangle \theta$ and  $\theta^{\prime}
v_i \partial_i \theta$. Their inclusion is manifested by the
multiplicative renormalization of the bare parameters $g_0, u_0$,
and $\nu_0$, and the velocity field ${\bf v}$ in the action
functional (\ref{action1}):
\begin{equation}
\nu_0=\nu Z_{\nu},\,\,\, g_0=g \mu^{2\varepsilon+\eta}
Z_g,\,\,\,u_0=u\mu^{\eta} Z_u,\,\,\, {\bf v}\rightarrow Z_v {\bf v}.
\label{zetka}
\end{equation}
Here, the dimensionless parameters $g,u$,and $\nu$ are the
renormalized counterparts of the corresponding bare ones, $\mu$ is
the renormalization mass (a scale setting parameter), and
$Z_i=Z_i(g,u,\alpha), i=\nu,g,u,v$ are renormalization constants.

The renormalized action functional has the following form
\begin{eqnarray}
%\hspace{-1cm}
\fl S(\Phi)=&-&\frac{1}{2} \int dt_1\,d^d{\bf x_1}\,dt_2\,d^d{\bf
x_2} v_i(t_1,{\bf x_1}) [D_{ij}^v(t_1,{\bf
x_1};t_2,{\bf x_2})]^{-1} v_j(t_2,{\bf x_2}) \label{actionRen} \\
&+& \int dt\,d^d{\bf x}\,\, \theta^{\prime}\left[-\partial_t \theta
- Z_2 v_i\partial_i\theta+\nu Z_1 \triangle\theta \right],\nonumber
\end{eqnarray}
where the correlator $D_{ij}^v$ is written in renormalized
parameters (in wave-number-frequency representation)
\begin{equation}
\tilde{D}_{ij}^v(\omega, k) = \frac{\left(P_{ij}({\bf k}) + \alpha
Q_{ij}({\bf k}) \right) g \nu^3 \mu^{2\varepsilon+\eta}
k^{4-d-2\varepsilon-\eta}}{(i\omega+u \nu \mu^{\eta}
k^{2-\eta})(-i\omega+u \nu \mu^{\eta}
k^{2-\eta})}.\label{corrveloRen}
\end{equation}
By comparison of the renormalized action (\ref{actionRen}) with
definitions of the renormalization constants $Z_i$, $i=g,u,\nu$
(\ref{zetka}) we are coming to the relations among  them:
\begin{equation}
Z_{\nu}=Z_1,\,\,\,Z_u=Z_1^{-1},\,\,\,Z_g=Z_2^2 Z_1^{-3},\,\,\,
Z_v=Z_2. \label{zetka1}
\end{equation}
The second and the third relations are consequences of the absence
of the renormalization of the term with $D^v$ in renormalized action
(\ref{actionRen}).

The issue of interest are especially multiplicatively renormalizable
equal-time two-point quantities $G(r)$ (see, e.g.,
\cite{Antonov00}). The example of such quantity are the equal-time
structure functions
\begin{equation}
S_{n}(r)\equiv\langle[\theta(t,{\bf x})-\theta(t,{\bf
x'})]^{n}\rangle \label{struc}
\end{equation}
in the inertial range, specified by the inequalities $l\sim
1/\Lambda <<r<<L=1/m$ ($l$ is an internal length). The infrared (IR)
scaling behavior of the function $G(r)$ (for $r/l\gg 1$ and any
fixed $r/L$)
\begin{equation}
G(r)\simeq \nu_0^{d^{\omega}_G} l^{-d_G} (r/l)^{-\Delta_G} R(r/L)
\label{frscaling}
\end{equation}
is related to the existence of IR stable fixed points of the RG
equations (see next section). In (\ref{frscaling}) $d^{\omega}_G$
and $d_G$ are corresponding canonical dimensions of the function
$G$, $R(r/L)$ is so-called scaling function which cannot be
determined by RG equation (see, e.g., \cite{Vasiliev}), and
$\Delta_G$ is the critical dimension defined as
\begin{equation}
\Delta_G=d_G^k+\Delta_{\omega} d_G^{\omega} + \gamma_G^*.
\end{equation}
Here $\gamma_G^*$ is the fixed point value of the anomalous
dimension $\gamma_G\equiv \mu \partial_{\mu} \ln Z_G$, where $Z_G$
is renormalization constant of multiplicatively renormalizable
quantity $G$, i.e., $G=Z_G G^R$ \cite{Antonov00}, and
$\Delta_{\omega}=2-\gamma_{\nu}^*$ is the critical dimension of
frequency with $\gamma_{\nu}=\gamma_1$ which is defined further in
the text (for more details see, e.g., \cite{Antonov99,Antonov00}).

On the other hand, the small $r/L$ behavior of the scaling function
$R(r/L)$ can be studied using the Wilson OPE \cite{Vasiliev}. It
shows that, in the limit $r/L\to 0$, the function $R(r/L)$ have the
following asymptotic form
\begin{equation}
R(r/L) = \sum_{F} C_{F}(r/L)\, (r/L)^{\Delta_F}, \label{ope}
\end{equation}
where $C_{F}$ are coefficients regular in $r/L$. In general, the
summation is implied over certain  renormalized com\-po\-si\-te
operators $F$  with critical dimensions $\Delta_F$.
% the end of the new text

In present paper we shall study only the first stage of the RG
analysis, namely, the influence of compressibility of the velocity
field on the stability of possible scaling regimes of the model. The
influence of compressibility on the anomalous scaling (the second
stage of the RG analysis) will be studied in the subsequent paper.

In what follows we shall work with two-loop approximation. But the
calculation of higher-order corrections is more difficult in the
models with turbulent velocity field with finite correlation time
than  in the cases with $\delta$-correlations in time.
%the old text
%First of all, one has to calculate more relevant Feynman diagrams in
%the same order of perturbation theory (see below). Second, and more
%problematic distinction, is related to the fact that the diagrams
%for the finite correlated case involve two different dispersion
%laws, namely, $\omega \propto k^2$ for the scalar field and $\omega
%\propto k^{2-\eta}$ for the velocity field. It leads to complicated
%expressions for re\-nor\-ma\-li\-za\-tion  constants even in the
%simplest (one-loop) approximation \cite{Antonov99,Antonov00}.
% the beginning of the new text
It is related to the fact that the diagrams for the finite
correlated case involve two different dispersion laws, namely,
$\omega \propto k^2$ for the scalar field and $\omega \propto
k^{2-\eta}$ for the velocity field which complicates situation even
in the one-loop approximation \cite{Antonov99,Antonov00}.
% the end of the new text
But, as was discussed in \cite{Antonov99,Antonov00,AdAnHo02}, this
difficulty can be avoided by the calculation of all renormalization
constants in an arbitrary specific choice of the exponents
$\varepsilon$ and $\eta$ that guarantees UV finiteness of the
Feynman diagrams. From the calculational point of view the most
suitable choice is to put $\eta=0$ and leave $\varepsilon$
arbitrary. Thus, the knowledge of the renormalization constants  for
the special choice $\eta=0$ is sufficient to obtain all important
quantities as the $\gamma$-functions, $\beta$-functions, coordinates
of fixed points, and the critical dimensions. But such possibility
is not automatic in general. In the model under consideration, it is
the consequence of an analysis which shows that in the minimal
subtraction (MS) scheme, which is used in what follows, all needed
anomalous dimensions are independent of the exponents $\varepsilon$
and $\eta$ in the two-loop approximation. But in the three-loop
approximation this dependence can simply appear \cite{AdAnHo02}.

Now let us continue with re\-nor\-ma\-li\-za\-tion of the model. The
relation $S(\theta,\theta^{\prime},{\bf v},
e_0)=S^R(\theta,\theta^{\prime},{\bf v}, e, \mu)$, where $e_0$
stands for the complete set of bare parameters and $e$ stands for
renormalized one, leads to the relation $W(A, e_0)=W^R(A, e, \mu)$
for the ge\-ne\-ra\-ting functional of connected Green functions. By
application of the operator $\tilde{\cal{D}}_{\mu}\equiv\mu
\partial_{\mu}$ at fixed $e_0$ on both sides of the latest equation
one obtains the basic RG differential equation
\begin{equation}
{\cal{D}}_{RG} W^R(A,e,\mu)=0, \label{RGE}
\end{equation}
where ${\cal{D}}_{RG}$ represents operation $\tilde{\cal{D}}_{\mu}$
written in the renormalized variables. Its explicit form is
\begin{equation}
{\cal{D}}_{RG} = {\cal{D}}_{\mu} +
\beta_g(g,u)\partial_g+\beta_u(g,u)\partial_u-\gamma_{\nu}(g,u){\cal{D}}_{\nu},\label{RGoper}
\end{equation}
where we standardly denote ${\cal{D}}_x\equiv x\partial_x$ for any
variable $x$, and the RG functions (the $\beta$ and $\gamma$
functions) are given by well-known definitions and, in our case,
using the relations (\ref{zetka1}) for renormalization constants,
they have the  following form
\begin{eqnarray}
\gamma_{i}&\equiv& \tilde{\cal{D}}_{\mu} \ln Z_{i},\,\,\, i=1,2 \label{gammanu}\\
\beta_g&\equiv&\mu \partial_{\mu} g =g
(-2\varepsilon-\eta+3\gamma_1-2\gamma_2), \label{betag}\\
\beta_u&\equiv&\mu \partial_{\mu} u =
u(-\eta+\gamma_1).\label{betau}
\end{eqnarray}

The renormalization constants $Z_1$, and $Z_2$ are determined by the
requirement that one-particle irreducible Green functions $\langle
\theta^{\prime} \theta \rangle_{1-ir}$ and $\langle \theta^{\prime}
\theta  {\bf v}\rangle_{1-ir}$ must be UV finite when are written in
renormalized variables. In our case, it means that they have no
singularities in the limit $\varepsilon, \eta\rightarrow0$.

\begin{figure}[t]
     \vspace{1cm}
       \begin{center}
       \leavevmode
       \epsfxsize=8.5cm
       \epsffile{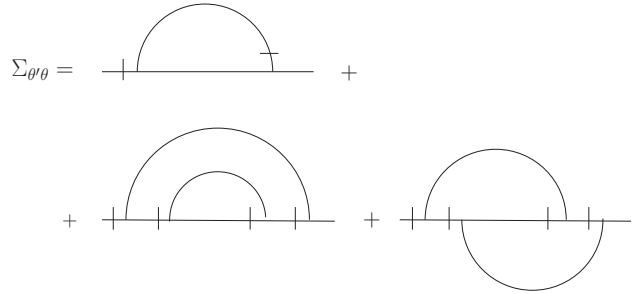}
   \end{center}
\vspace{-0.5cm} \caption{The one and two-loop diagrams which
contribute
\\ to the self-energy operator $\Sigma_{\theta^{\prime}\theta}$.
\label{fig3}}
\end{figure}

The one-particle irreducible Green function $\langle \theta^{\prime}
\theta\rangle_{1-ir}$ is related to the self-energy ope\-ra\-tor
$\Sigma_{\theta^{\prime}\theta}$ by the Dyson equation
\begin{equation}
\langle \theta^{\prime}\theta \rangle_{1-ir}=-i\omega+\nu_0 p^2 -
\Sigma_{\theta^{\prime}\theta}(\omega, p),\label{Dyson}
\end{equation}
where the self-energy operator $\Sigma_{\theta^{\prime}\theta}$ is
represented by the corresponding one-particle irreducible diagrams.
In the two loop approximation, it is defined by the diagrams which
are shown in figure \ref{fig3}.
\input epsf

\begin{figure}[t]
     \vspace{1cm}
       \begin{center}
       \leavevmode
       \epsfxsize=9cm
       \epsffile{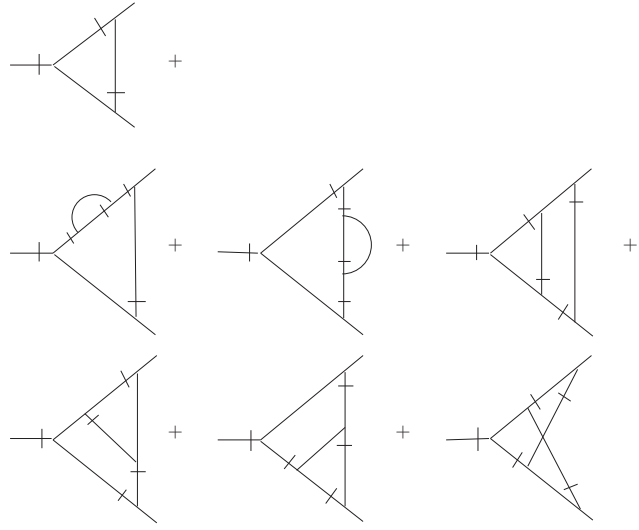}
   \end{center}
\vspace{-0.5cm} \caption{The one and two-loop diagrams which
contribute to the function ${\cal V}_i$. \label{fig4}}
\end{figure}

On the other hand, the renormalized function $\langle
\theta^{\prime} \theta {\bf v}\rangle_{1-ir}$ is defined as
\begin{equation}
\langle\theta^{\prime} \theta  v_i\rangle_{1-ir}= Z_2 V_i + {\cal
V}_i,\label{vrcholl}
\end{equation}
where the function ${\cal V}_i$ is defined by diagrams of figure
\ref{fig4} (in two-loop approximation).
\input epsf

Thus, $Z_{1}$, and $Z_2$ are found from the requirement that the UV
divergences are canceled in (\ref{Dyson}), and  (\ref{vrcholl})
after substitution $\nu_0=\nu Z_{\nu}=\nu Z_{1}$. This determines
$Z_{1}$, and $Z_2$ up to an UV finite contribution, which are fixed
by the choice of the renormalization scheme. In the MS scheme all
renormalization constants have the form: 1 + {\it poles in
$\varepsilon,\eta$ and their li\-near combinations}. As was already
mentioned, in our calculations we can put $\eta=0$. This possibility
essentially simplifies the eva\-lua\-tions of all quantities
\cite{Antonov99,Antonov00,AdAnHo02}.  The analytical expressions for
one-loop diagrams in figure \ref{fig3} and figure \ref{fig4} (in the
MS scheme) have the following form
\begin{eqnarray}
G_{1p}&=&-\frac{S_d}{(2\pi)^d}\frac{g \nu p^2}{4 u (1+u)^2}\frac{(1
+ u)(d - 1 + \alpha) - 2\alpha}{d}
\left(\frac{\mu}{m}\right)^{2\varepsilon}\frac{1}{\varepsilon},\\
G_{1v}&=& i \frac{S_d}{(2\pi)^d} \frac{g  p_j}{4 u
(1+u)^2}\frac{\alpha}{d}
\left(\frac{\mu}{m}\right)^{2\varepsilon}\frac{1}{\varepsilon},
\end{eqnarray}
where $G_{1p}$ is result for the one-loop diagram in figure
\ref{fig3}, and $G_{1v}$ is result for the one-loop diagram in
figure \ref{fig4}. Here, $S_d=2 \pi^{d/2}/\Gamma(d/2)$ denotes the
$d$-dimensional sphere. The two-loop expressions for the diagrams in
figure \ref{fig3} and figure \ref{fig4} are rather huge, therefore
we shall not present their explicit form separately but rather we
present complete expressions for renormalization constants $Z_1$,
and $Z_2$ which have the following structure
\begin{equation}
Z_i=\frac{g}{\varepsilon} A_i +
\frac{g^2}{\varepsilon}\left(\frac{1}{\varepsilon}B_i+C_i\right),\,\,\,i=1,2.\label{ZZZ}
\end{equation}

Now using the definition of the anomalous dimensions $\gamma_{1,2}$
in (\ref{gammanu}) we obtain
\begin{eqnarray}
\gamma_1&\equiv&\mu \partial_{\mu} \ln Z_1 = -2 (\bar{g} A_1 + 2 \bar{g}^2 C_1),\label{g1}\\
\gamma_2&\equiv&\mu \partial_{\mu} \ln Z_2 = -2 (\bar{g} A_2 + 2
\bar{g}^2 C_2), \label{g2}
\end{eqnarray}
where we denote $\bar{g}=g S_d/(2\pi)^d$. The one-loop contributions
$A_1$ and $A_2$ in (\ref{g1}) and (\ref{g2}) are defined as follows
\begin{eqnarray}
A_1&=& -\frac{1}{4 u (1+u)^2}\frac{(1 + u)(d - 1 + \alpha) - 2\alpha}{d}, \label{aa1}\\
A_2&=&\frac{\alpha}{4 d u (1+u)^2},\label{aa2}
\end{eqnarray}
and the two-loop contributions $C_1$ and $C_2$ have the form
\begin{eqnarray}
C_1&=&\frac{1}{16 d^2 u^2 (1+u)^3}\left(C_{10} +\alpha C_{11} +
\alpha^2 C_{12} \right), \label{cc1}\\
C_2&=&\frac{1}{32 d^3 u^2 (1+u)^6}\left(\alpha C_{21} + \alpha^2
C_{22}\right),\label{cc2}
\end{eqnarray}
where
\begin{eqnarray}
&&\hspace{-2.3cm} C_{10}=\frac{(d-1)(d+u) H_2}{(d+2)(1+u)^2},\\
&&\hspace{-2.3cm} C_{11}=\frac{d-1}{1+u}-\frac{u (d-1)(2 + u)(2
(u-2) u + d (2 + 3
u))}{d(1+u)^3} H_0 \nonumber \\
&& \hspace{-1.3cm}+\frac{(d-1) (4 (-2 + u) u + d^2 (-2 + 3 u^2 (2 +
u)) + 2 d (2 + u (5 - 5 u + u^3)))}{d^2(1+u)^3} H_1,\\
&&\hspace{-2.3cm} C_{12}=\frac{3u-1}{(1+u)^2} + \frac{u H_0}{d(1+u)^4}\Big(2 d^2 u (1+u)^2
-(u-3) (u-1) u (2 + u)\nonumber \\
&&\hspace{2.5cm}   + d (2 + u) (1 + u (-2 + (u-6) u)) \Big)
\nonumber \\ && \hspace{-1.3cm} +\frac{H_1}{d^2(1+u)^4} \Big( 2
(u-3) (u-1) u - 2 d^3 u^2 (1 + u)^2 \nonumber \\
&& \hspace{1.3cm} - d^2 (-1 + u (5 + u (2 + u) (3 + (u-6) u)))
\nonumber \\ && \hspace{1.3cm} + d (-2 + u (1 + u (22 + (u-4) u (2 +
u))))\Big), \\
&&\hspace{-2.3cm} C_{21}= -(d-1) d (1 + u)^2 -(d-1) d u (3 + 5 u + 2
u^2) H_0 \nonumber \\
&& \hspace{-1.3cm}+
(d-1) (1 + 2 u) (2 + d u (2 + u)) H_1  + \frac{2 (d-1) d (u-1)}{d+2}H_2, \\
&&\hspace{-2.3cm} C_{22}= -4 d (1 + u)\nonumber \\
&& \hspace{-1.3cm} + u (-2 (2 + u) + d (5 - 2 u (1 + u) + d (-1 + 2 u (1 + u)))) H_0 \nonumber \\
&& \hspace{-1.3cm} + \frac{2 (2 (1 + u) - d^3 u^2 (1 + u)^2 + d (-3
+ u + 5 u^2 + u^3) + d^2 (1 - u + u^3 + u^4))}{d (1 + u)}H_1\nonumber \\
&& \hspace{-1.3cm} + \frac{2 d ((u-1)^2 + d (1 + 2 u - u^2))}{(2 +
d) (1 + u)}H_2,
\end{eqnarray}
where we have used the following notation
\begin{equation}
H_i={_2F_1}\left[1,1;i+\frac{d}{2};\frac{1}{(1+u)^2}\right],\,\,\,
i=0,1,2
\end{equation}
for the corresponding hypergeometric function
${_2F_1}[a,b;c;z]=1+\frac{a\,
b}{c\cdot1}z+\frac{a(a+1)b(b+1)}{c(c+1)\cdot1\cdot2}z^{2}+\ldots$.
%the beginning of the added text
The functions $B_i, i=1,2$ which are introduced in (\ref{ZZZ}) are
not important in what follows, therefore we shall not define them
explicitly.
% the end of the added text

\section{Fixed points and scaling regimes}\label{sec5}

Possible scaling regimes of a renormalizable model are directly
given by the IR stable fixed points of the corresponding system of
RG equations \cite{Vasiliev,ZinnJustin}. The coordinates of the
fixed point of the RG equations are defined by $\beta$-functions,
namely, by requirement of their vanishing. In our model the
coordinates $g_*, u_*$ of the fixed points are found from the system
of two equations
\begin{equation}
\beta_g(g_*,u_*)=\beta_u(g_*,u_*)=0.
\end{equation}
The beta functions $\beta_g$ and $\beta_u$ are defined in
(\ref{betag}) and (\ref{betau}). The IR asymptotic behavior is
governed by the IR stable fixed point which is given by the positive
eigenvalues of the matrix $\Omega$ of the first derivatives:
\begin{equation}
\Omega_{ij}=\left(\begin{array}{cc}\partial \beta_g/\partial g &
\partial \beta_g/\partial u \\ \partial \beta_u/\partial g & \partial \beta_u/\partial u
\end{array}
\right).
\end{equation}

The influence of compressibility on the scaling regimes of the
present model in one-loop approximation was investigated in
\cite{Antonov00}. We are interested in the answer on the following
question: how can the two-loop approximation change the picture of
the scaling regimes discussed in \cite{Antonov00}?

In what follows, we shall try to study possible scaling regimes in
detail. First of all, we shall investigate the rapid-change limit:
$u\rightarrow\infty$. In this regime, it is necessary to make
transformation to new variables, namely, $w\equiv1/u$, and
$g^{\prime}\equiv g/u^2$, with the corresponding changes in the
$\beta$ functions:
\begin{eqnarray}
\beta_{g^{\prime}}&=&g^{\prime}
(-2\varepsilon+\eta+\gamma_1-2\gamma_2), \label{betag1}\\
\beta_w &=& w(\eta-\gamma_1).\label{betau1}
\end{eqnarray}
It is well-known that in the rapid change model the higher than
one-loop corrections to the self-energy operator are equal to zero.
On the other hand, the renormalization of the velocity field is
absent at all as a consequence of the fact that $Z_2=1$ at all
orders of the perturbation theory. It can be also seen directly by
the corresponding manipulations with our $\gamma$-functions
(\ref{g1}) and (\ref{g2}). Therefore, we are coming to the one-loop
results of \cite{Antonov00} (in the rapid-change model limit),
namely
\begin{equation}
\gamma_1=\bar{g}^{\prime}\frac{d-1+\alpha}{2d},\,\,\,\gamma_2=0,
\label{gamma10}
\end{equation}
where again $\bar{g}^{\prime}=g^{\prime} S_d/(2\pi)^d$.

In this regime we have two fixed points denoted as FPI and FPII in
\cite{Antonov99,Antonov00}. The first of them is trivial one
\begin{equation}
\mathrm{FPI}:\,\,\,\, w_*=g_*^{\prime}=0,
\end{equation}
with $\gamma_{1}^*=0$, and diagonal matrix $\Omega$ with eigenvalues
(diagonal elements)
\begin{equation}
\Omega_1=\eta,\,\,\,\,\,\Omega_2=\eta-2\varepsilon.
\end{equation}
Thus, this fixed point is IR stable when $\eta>0$, and, at the same
time, $\eta>2\varepsilon$. The second point is defined as
\begin{equation}
\mathrm{FPII}:\,\,\,\,w_*=0,\,\,\,
\bar{g}_*^{\prime}=\frac{2d}{d-1+\alpha}(2\varepsilon-\eta),
\end{equation}
with exact one loop result $\gamma_{1}^*=2\varepsilon-\eta$. The
corresponding $\Omega$  matrix is triangular with diagonal elements
(eigenvalues):
\begin{equation}
\Omega_1=2(\eta-\varepsilon),\,\,\,\,\Omega_2=2\varepsilon-\eta.
\end{equation}
It means that this kind of the fixed point is IR stable when
$\eta<2\varepsilon$ together with $\eta>\varepsilon$.

The second special case of the present model is so-called "frozen
regime" with the frozen velocity field. It is obtained from our
model in the limit $u\rightarrow0$. To consider this transition, it
is again appropriate to change the variable $g$  to the new variable
$g^{\prime\prime} \equiv g/u$ \cite{Antonov99}. Then the $\beta$
functions are transform to the following ones:
\begin{eqnarray}
\beta_{g^{\prime\prime}}&=&g^{\prime\prime}
(-2\varepsilon+2\gamma_1-2\gamma_2), \label{betag2}\\
\beta_u &=& u(-\eta+\gamma_1),\label{betau2}
\end{eqnarray}
with unchanged $\beta$ function for parameter $u$. In this notation,
the anomalous dimensions $\gamma_{1,2}$ have the form
\begin{eqnarray}
\gamma_1&=& -2 (\bar{g}^{\prime\prime} A_1^{\prime\prime} +
2 \bar{g}^{\prime\prime2} C_1^{\prime\prime}),\label{g1u0}\\
\gamma_2&=& -2 (\bar{g}^{\prime\prime} A_2^{\prime\prime} + 2
\bar{g}^{\prime\prime2} C_2^{\prime\prime}), \label{g2u0}
\end{eqnarray}
where, as obvious,  $\bar{g}^{\prime\prime}=g^{\prime\prime}
S_d/(2\pi)^d$, and the one-loop contributions are now given as
\begin{eqnarray}
A_1^{\prime\prime}&=& -\frac{d - 1 - \alpha}{4d},  \\
A_2^{\prime\prime}&=&\frac{\alpha}{4d},
\end{eqnarray}
and the two-loop contributions $C_1^{\prime\prime}$ and
$C_2^{\prime\prime}$ are now
\begin{eqnarray}
C_1^{\prime\prime}&=&\frac{1}{16 d^2}\left(C_{10}^{\prime\prime}
+\alpha C_{11}^{\prime\prime} +
\alpha^2 C_{12}^{\prime\prime} \right), \\
C_2^{\prime\prime}&=&\frac{1}{32 d^3}\left(\alpha
C_{21}^{\prime\prime} + \alpha^2 C_{22}^{\prime\prime}\right),
\end{eqnarray}
with
\begin{eqnarray}
&&\hspace{-2.3cm} C_{10}^{\prime\prime}=\frac{(d-1)d }{(d+2)}H_{02}=d-1,\\
&&\hspace{-2.3cm} C_{11}^{\prime\prime}=(d-1)\left(1 -\frac{2(d - 2 )}{d} H_{01}\right)=1-d,\\
&&\hspace{-2.3cm} C_{12}^{\prime\prime}=-1+\frac{d  - 2}{d}H_{01}=0, \\
&&\hspace{-2.3cm} C_{21}^{\prime\prime}= 2(d-1)\left( H_{01} - \frac{d}{d+2}H_{02}\right)=\frac{4(d-1)}{d-2}, \\
&&\hspace{-2.3cm} C_{22}^{\prime\prime}= \frac{2 (d-1)(d-2)}{d
}H_{01}+ \frac{2 d (1 + d)}{(2 + d)}H_{02}=4d,
\end{eqnarray}
where we denote
\begin{equation}
H_{0i}={_2F_1}\left[1,1;i+\frac{d}{2};1\right]=\frac{d-2+2i}{d-4+2i},\,\,\,
i\geq1.
\end{equation}
The system of $\beta$ functions (\ref{betag2}) and (\ref{betau2})
exhibits two fixed points, denoted as FPIII and FPIV in
\cite{Antonov99}. They are related to the corresponding two scaling
regimes. One of them is trivial,
\begin{equation}
\mathrm{FPIII}:\,\,\,\, u_*=g_*^{\prime\prime}=0,
\end{equation}
with $\gamma_{1}^*=\gamma_2^*=0$. The eigenvalues of the
corresponding matrix $\Omega$, which is diagonal in this case, are
\begin{equation}
\Omega_1=-2\varepsilon,\,\,\,\,\Omega_2=-\eta.
\end{equation}
Thus, this regime is IR stable only if both parameters
$\varepsilon$, and $\eta$ are negative simultaneously. The second,
non-trivial, point is
\begin{equation}
\mathrm{FPIV}:\,\,\,\, u_*=0,\,\,\,\,
\bar{g}_*^{\prime\prime}=-\frac{\varepsilon}{2
(A_1^{\prime\prime}-A_2^{\prime\prime})}-\frac{C_1^{\prime\prime}-C_2^{\prime\prime}}{2
(A_1^{\prime\prime}-A_2^{\prime\prime})^3} \varepsilon^2,
\end{equation}
with exact one-loop relation $\gamma_1^*=\gamma_2^*+ \varepsilon$.
After substitution of the corresponding quantities one obtains the
following expression for the coordinates of the fixed point
\begin{equation}
\hspace{-1.8cm}u_*=0,\,\,\,\, \bar{g}_*^{\prime\prime}=\frac{2 d
\varepsilon}{d-1}\left\{1+\frac{\varepsilon}{(d-1)^2}
\left[(d-1)\left(1-\alpha\left(1+\frac{2}{d(d-2)}\right)\right)-2\alpha^2\right]\right\}.
\end{equation}
The eigenvalues of the matrix $\Omega$ (taken at the fixed point)
are
\begin{equation}
\Omega_1=2\varepsilon \left(1-
\frac{C_1^{\prime\prime}-C_2^{\prime\prime}}{(A_1^{\prime\prime}-A_2^{\prime\prime})^2}\varepsilon\right),\,\,\,
\Omega_2=\varepsilon-\eta+\gamma_2^*.
\end{equation}
After corresponding substitutions one has
\begin{eqnarray}
&&\hspace{-2cm}\Omega_1=2
\varepsilon\left\{1-\frac{\varepsilon}{(d-1)^2}
\left[(d-1)\left(1-\alpha\left(1+\frac{2}{d(d-2)}\right)\right)-2\alpha^2\right]\right\},
\\ && \hspace{-2cm}\Omega_2=\varepsilon-\eta + \frac{\alpha\, \varepsilon}{d-1}\left[-1+ \varepsilon
\frac{2 \alpha^2 (d-2) d - (d-1)(d^2-2)-\alpha (2+(d-3)
d^2)}{(d(d-1)^2(d-2))} \right].
\end{eqnarray}
The conditions $\bar{g}_*^{\prime\prime}>0, \Omega_{1}>0$, and
$\Omega_2>0$ for the IR stable fixed point lead to the restrictions
on the values of the parameters $\varepsilon$ and $\eta$. First,
suppose that $\varepsilon<0$. Then from the conditions
$\bar{g}_*^{\prime\prime}>0$, and  $\Omega_{1}>0$ one has the
following restrictions which must be fulfilled simultaneously
\begin{equation}
1+\varepsilon D<0,\,\,\,1-\varepsilon D<0,\label{cond1}
\end{equation}
but they cannot be fulfilled at the same time. Thus, our first
condition is $\varepsilon>0$. In (\ref{cond1}) $D$ is given as
\begin{equation}
D=\frac{1}{(d-1)^2}
\left[(d-1)\left(1-\alpha\left(1+\frac{2}{d(d-2)}\right)\right)-2\alpha^2\right].
\end{equation}
To have $\bar{g}_*^{\prime\prime}>0$, and  $\Omega_{1}>0$ together
with $\varepsilon>0$, the following inequalities must be held
\begin{equation}
-1< \varepsilon D < 1, \label{cond2}
\end{equation}
which restricts the value of $\varepsilon$ as a function of the
parameter $\alpha$, and the dimension of the space $d$. In the
incompressible case ($\alpha=0$) the condition (\ref{cond2}) is
reduced into the simple inequality
\begin{equation}
\varepsilon<d-1.
\end{equation}
In the general case, for each value of $d$, there exists a
"critical" value of $\alpha$  in which $D=0$. We denote it as
$\alpha_c$. In this situation $\varepsilon$ can be arbitrary, i.e.,
the condition (\ref{cond2}) is fulfilled automatically. The value of
$\alpha_c$ is defined as follows
\begin{equation}
\hspace{-1.5cm}\alpha_c=\frac{2-4 d+3 d^2-d^3+(4-16 d-4 d^2+36
d^3-23 d^4+2 d^5+d^6)^{1/2}}{4 d (d-2)}.
\end{equation}
For example, for $d=3$ its value is
$\alpha_c=\frac{\sqrt{61}-5}{6}\simeq0.468$. Therefore, in the
compressible model, the situation is a little bit more complicated
as a result of a competition between incompressible and compressible
terms within two-loop approximation which leads to the existence of
$\alpha_c$. How does it work? The answer is the following. If we
continuously increase the value of the parameter $\alpha$, the
region of stability of the fixed point defined by the inequalities
(\ref{cond2}) increases too. This restriction vanishes completely
when $\alpha$ reaches the "critical" value $\alpha_c$. In this
rather specific situation the two-loop influence on the region of
stability of fixed point defined by condition (\ref{cond2}) is
exactly zero. Then, if the value of parameter $\alpha$ increases
further, the condition (\ref{cond2}) appears again, and restriction
on $\varepsilon$ becomes stronger when $\alpha$ tends, in principle,
to infinity. In this limit $\varepsilon \rightarrow 0$. On the other
hand, it must be stressed that in our model only relatively small
values of $\alpha$ are allowed ($\alpha \ll 1$). It corresponds to
small fluctuations of the density $\rho$ in the system which is
supposed in our investigation. In other words, it is supposed that
the stochastic component of the velocity field of the fluid is much
smaller than the velocity of the sound in the system (the Mach
number $Ma\ll 1$).

The last condition on the stability of the IR fixed point is found
from the requirement to have $\Omega_2>0$. It reads
\begin{equation}
\hspace{-1cm}\eta< \varepsilon + \frac{\alpha\,
\varepsilon}{d-1}\left[-1+ \varepsilon \frac{2 \alpha^2 (d-2) d -
(d-1)(d^2-2)-\alpha (2+(d-3) d^2)}{(d(d-1)^2(d-2))} \right].
\end{equation}
In the incompressible case it is reduced into the simple condition
\begin{equation}
\eta<\varepsilon,
\end{equation}
which is held at each order of the perturbation theory.

In the end, let us consider the most interesting scaling regime with
finite value of the fixed point for variable $u$. The coordinates of
the fixed point is now defined by the requirement of vanishing of
the $\beta$ functions which are given in (\ref{betag}) and
(\ref{betau}). The fixed point value for $\bar{g}=g S_d/(2\pi)^d$ is
given as
\begin{equation}
\mathrm{FPV}:\,\,\,\, \bar{g}_*=-\frac{\varepsilon}{2(A_{1}-A_{2})}-
\frac{C_{1}-C_{2}}{2 (A_{1}-A_{2})^3} \varepsilon^2,
\end{equation}
where the functions $A_1, A_2, C_1$, and $C_2$ are given in
(\ref{aa1})-(\ref{cc2}), and where the parameter $u$ is taken at its
fixed point value $u_*$ which is given implicitly by the equation
\begin{equation}
-\eta+\gamma_1^*(u_*)=0.
\end{equation}
Using the exact relations
\begin{equation}
\gamma_1^*=\eta,\,\,\,\,\,\gamma_2^*=\eta-\varepsilon
\end{equation}
the expression for the fixed point value of $\bar{g}$ can be
rewritten as a series (expansion) of the parameter $\eta$ or a
linear combination of $\eta$ and $\varepsilon$. For example, in
\cite{Antonov00}, where the problem was analyzed in one-loop
approximation, it was expressed as a function of $2\varepsilon-\eta$
(in our notation). In the framework of one-loop approximation it
allows one to have linear dependence of $\bar{g}_*$ on the fixed
point value of the parameter $u$. Together with another choice of
the linear combination of $\eta$ and $\varepsilon$, namely
$\eta-\varepsilon$ it leads also to the simple expression for the
fixed point value of $u$. Thus, the coordinates of the fixed point
in one-loop approximation are \cite{Antonov00}
\begin{equation}
\bar{g}_*=\frac{2 d
(1+u_*)}{d-1+\alpha}(2\varepsilon-\eta),\,\,\,\,\,\,
u_*=-1+\frac{\alpha}{d-1+\alpha}\frac{\eta-2\varepsilon}{\eta-\varepsilon}.
\end{equation}
It allows, together with the requirement of the positive eigenvalues
of the corresponding matrix of the first derivatives $\Omega$, to
find simple conditions for the IR stable fixed point. They are
defined by inequalities $\varepsilon>0, \varepsilon>\eta$, and
$\eta>\varepsilon\frac{d-1-\alpha}{d-1}$ \cite{Antonov00}.

The situation is essentially more complicated when we are working in
two-loop approximation. It is given by the fact that now we have
nonlinear dependence of $\bar{g}$ on the parameters $\eta$ and
$\varepsilon$, and the expression for the fixed point value of $u$
is now given only implicitly in rather complicated expression
containing hypergeometric functions. Another complication, which
defends to analyze the problem in general, is related to the fact
that contrary to the incompressible case when one has additional
condition, namely $\eta=\varepsilon$, no such condition exists in
compressible case under consideration. As a result, the analysis of
the IR stability of the general case of the present model have to be
done individually for concrete situation. It is rather cumbersome
and it will be done in the subsequent work.

In what follows, let us only give the general analysis of the most
interesting case when one suppose the relation $\eta=\varepsilon$.
In this situation from the definition of the $\beta$ functions given
in (\ref{betag}) and (\ref{betau}) one obtains the condition
\begin{equation}
\gamma_2^*=0.
\end{equation}
Thus, in this case, the coordinates of the fixed points are given as
\begin{eqnarray}
&&\bar{g}_*=-\frac{\varepsilon}{2 A_{1}}- \frac{C_{1}}{2 A_{1}^3}
\varepsilon^2, \\
&& A_2(u_*) + 2 \bar{g}_* C_2(u_*)=0, \label{impu}
\end{eqnarray}
but even in this situation the fixed point value of $u$ is defined
by complicated implicit equation (\ref{impu}) and its exhausted
analysis must be discussed separately.

\section{Conclusions}\label{sec6}

We have studied the influence of compressibility on the possible IR
scaling regimes of the model of a passive scalar advected by a
Gaussian velocity field with finite time correlations by means of
the field theoretic RG technique. The possible scaling regimes are
directly connected to the existence of IR stable fixed points of the
RG equations. The dependence of the fixed points on the parameter of
compressibility and their IR stability is discussed.
% the beginning of the added text
The most attention is paid to the frozen limit of the model where
inequalities which define the stable IR scaling regimes are found
analytically. The existence of a "critical" value $\alpha_c$ of the
parameter of compressibility $\alpha$ at which one of the two-loop
conditions is canceled as a result of the competition between
compressible and incompressible terms is discussed in detail. The
main conclusion is that for the small value of parameter $\alpha$
the region of stability is not restricted considerably.
% the end of the added text
It is also shown that the most general case with finite time
correlations of the velocity field is more complicated within
two-loop approximation and have to be consider separately once more.

\ack %\section*{Acknowledgement}
The work was supported in part by VEGA grant
% 3211 and
6193 of Slovak Academy of Sciences, by Science and Technology
Assistance Agency under contract No. APVT-51-027904.

\end{document}